\documentstyle[epsf,proceedings]{crckapb}
\begin{opening}
\title{DIFFERENTIAL STUDIES OF CEPHEIDS IN THE MAGELLANIC CLOUDS}
\author{J.P. Beaulieu}
\institute{Kapteyn Laboratorium, postbus 800, 9700AV Groningen}
\author{D.D. Sasselov}
\institute{Harvard-Smithsonian Center for Astrophysics, Cambridge, MA 02138, USA}
\end{opening}
\begin{document}

\section{Introduction}
EROS (Aubourg et al., 1993) is a French 
collaboration to search for baryonic dark matter
in the Galactic Halo by microlensing on stars of the Magellanic Clouds. 
Between 1991 and 1994 about 15000
CCD images were taken with two broad bandpass filters $B_E$ and $R_E$ of an area of 
1 x 0.4 degrees centered  on the bar of the LMC. Another field of 1 x 0.4 degrees had been
chosen in the SMC between 1993-1995 and about 6000 images where taken. 
Lightcurves of 270 000 stars exist with up to 48 points per 
night ! 
We have searched systematically the EROS data base for variable stars using the modified 
periodogram technique (Grison 1994) and the AoV method (Schwarzenberg-Czerny 1989).
We delineated the instability strip in the color magnitude diagram in order to 
exclude eclipsing binaries. The remaining 97 variable stars in the LMC (Beaulieu et al. 1995) 
and 450 in the SMC (Beaulieu \& Sasselov 1997) 
form our sample of Cepheids. Of them, one in the LMC, and 11 in the SMC (Beaulieu et al.,
1997) are "beat Cepheids" [BCs] $-$ pulsating simultaneously in two modes. The SMC search for 
beat Cepheids is preliminary: we plan to apply a better signal processing analysis 
to this data set.

\section{Classical and s-Cepheids}
A Cepheid envelope is an acoustic cavity in which an infinity of modes of pulsation 
exist. However, very few of them contribute to the dynamics of the system; the unstable modes
and the marginally stable modes coupled by resonances to unstable modes.
Resonances are known to play an important role in shaping the light curves. 

Usually, a Fourier decomposition of the light curve of the form 
$X = X_0 + \sum_{k=i}^N X_k \cos (k \omega_k t + \Phi_k)$ is computed. The phase difference
$\Phi_{k1}=\Phi_k - k \Phi_1$ and the amplitude ratio $R_{k1} = X_k \ X_1$ are calculated 
and used to look for resonances between pulsational modes.
In Figure 1 we show $R_{21}$ and $\Phi_{21}$ as a function of period for the Galactic
short period Cepheids (Antonello et al., 1996), LMC Cepheids and SMC Cepheids observed by EROS
(Beaulieu \& Sasselov 1997 and references therein). The so-called classical Cepheids are plotted
as crosses, and the s-Cepheids $-$ with more sinusoidal curves of lower amplitude,
are plotted as
filled diamonds. We follow the classification based on the $R_{21}-P$ plane proposed by 
Antonello et al. (1986). These authors classified as s-Cepheids the stars lying in the lower
part of the $R_{21}-P$ diagram. This morphological classification is mirrored by a dichotomy in the
period-luminosity plane (Figure 2). Therefore we conclude that the difference is a consequence 
of different pulsational modes: a fundamental radial mode for classical Cepheids (F), 
and a first overtone radial mode for s-Cepheids (1OT).
\begin{figure}
\epsfxsize=12cm
\epsfysize=12cm
\centerline{\epsffile{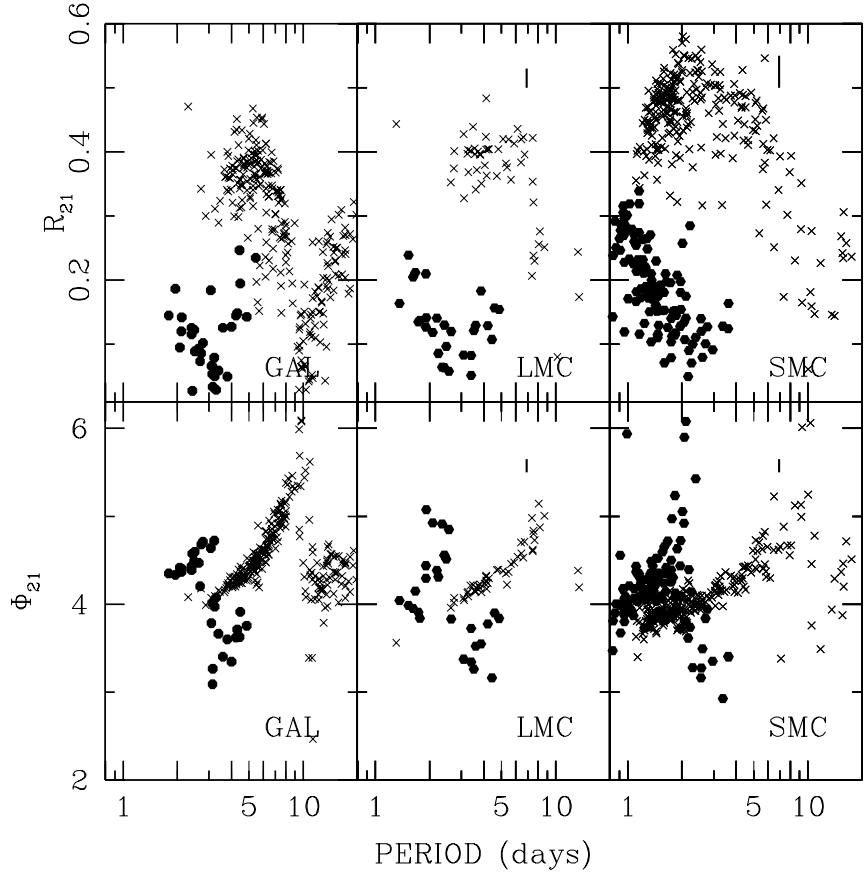}}
\caption[]{ The $R_{21}-P$, $\Phi_{21}-P$ planes for 
short period Galactic Cepheids
and EROS LMC and SMC Cepheids. S-Cepheids are plotted as filled diamonds, 
classical Cepheids are plotted as crosses.  
Typical error bars are indicated in the upper part of each subfigure.}
\end{figure}

\begin{figure}
\epsfxsize=12cm
\epsfysize=12cm
\centerline{\epsffile{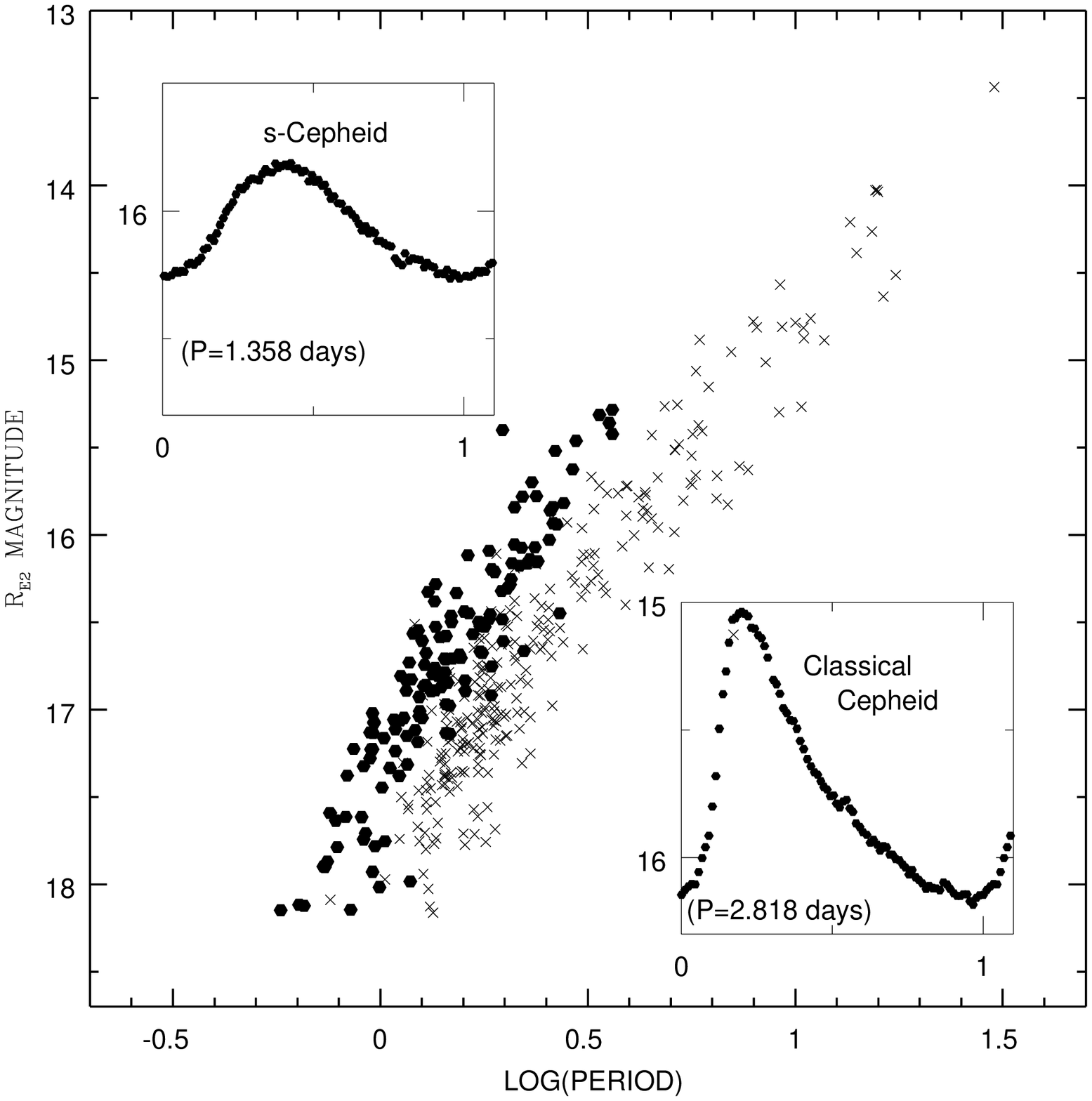}}
\caption[]{ The Period-Luminosity relation for SMC Cepheids from the EROS database.
 Classical Cepheids (fundamental mode pulsators) are plotted as crosses, and s-Cepheids 
(overtone pulsators)
are plotted as dots. Two examples of classical (lower right) and s-Cepheids (upper left) 
light curves are shown in the figure. The same scale for the amplitude has been 
adopted. 
The two types of Cepheids can be isolated on morphological basis, and follow different 
PL relations. }
\end{figure}

\subsection{Classical Cepheids}
In the three galaxies, we observed the Hertzsprung progression of the changing form of 
Cepheid light curves due to a 2:1 resonance between the fundamental and the second overtone.
This resonance takes place at $10\pm0.5$ days in our Galaxy. Using data obtained by the 
MACHO microlensing survey in the LMC in Figure 1 of Welch et al. (1996), the
resonance takes place between 10-11 days. From EROS SMC observation, the resonance in the 
SMC takes place in the range 10.5-11.5 days.
Cepheids in the three galaxies follow the same sequence before the resonance.
However, Cepheids of shorter and shorter periods are found in the SMC and LMC, compared to
our Galaxy; the shortest period Cepheids are observed in the metal-poorest SMC sample.
The metallicity affects the extension of the blue loops of the evolutionary tracks.
For high metallicity, the blue loops for low mass stars do not cross the instability strip,
whereas they do for low metallicity.

\subsection{s-Cepheids}
The s-Cepheids are observed in the three galaxies. They present the same behaviour in the
Fourier plane: a 
rise of $\Phi_{21}$, followed by a sharp drop. This is mirrored by a minimum in $R_{21}$.
It has been proposed that this drop is a signature of a 2:1 resonance between the first and
the fourth overtone. The drop takes place at $3.2 \pm 0.2$ days in our Galaxy, $2.7 \pm 0.2$ days
in the LMC, and $2.2 \pm 0.2 $ days in the SMC.

\section{Beat Cepheids}
So far 73 beat Cepheids have been found in the LMC by MACHO (Welch et al. 1996 
and references therein).
We report the discovery of 11 BCs in the SMC (Beaulieu et al., 1997), and 
14 are known in our Galaxy (Pardo \& Poretti 1997 and references therein). 
Their period ratios are plotted in Figure 2. On the upper figure are shown the BCs
pulsating both in the first and second overtone (1OT/2OT) and on the lower figure $-$ those 
pulsating in the fundamental and the first overtone.
The SMC 1OT/2OTs are very similar to the LMC ones while the SMC F/1OTs have period ratios
systematically higher than the LMC ones by $\sim 0.01$, which are systematically higher
than the Galactic ones by $\sim 0.01$.
In the case of a F/1OT pulsator, the fundamental mode penetrates deeper in the envelope 
of the star than the
first overtone. Therefore, the changed period ratio is mainly due to the effect 
of opacity on the fundamental
period. The case of 1OT/2OT is different since these two modes are more concentrated 
to the surface.
 We can crudely consider that an upward change in metallicity is an increase
of the opacity in the opacity bump due to metals found at depths around
100 000 K. The driving of the pulsation comes from the helium ionization zone, while
 the metals opacity bump  affects the acoustic properties of the envelope.
\begin{figure}
\epsfxsize=9cm
\epsfysize=9cm
\centerline{\epsffile{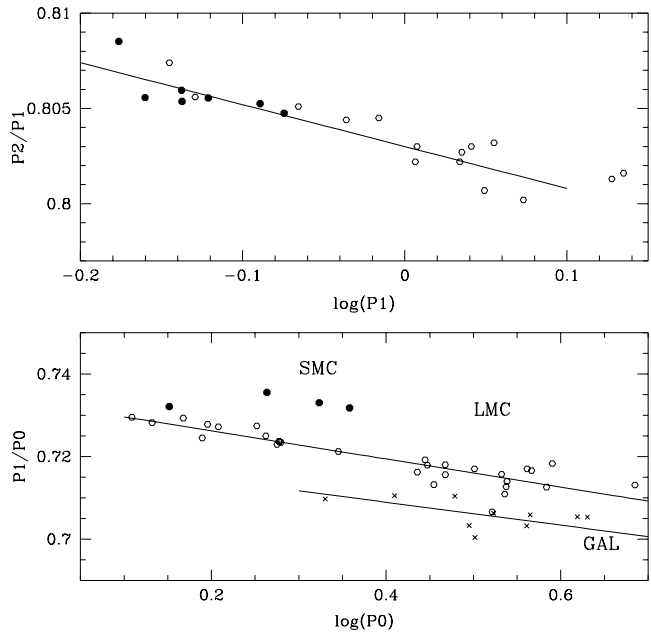}}
\caption[]{ Period ratio versus period for Galactic (crosses), LMC (circles) and
 SMC (dots) beat Cepheids.}
\end{figure}
With these two kinds of beat Cepheids, plus the two resonance constraints for the classical
Cepheids and the s-Cepheids observed at different metallicities, we are probing different 
depths in the Cepheid envelopes, and drawing new strong constraints (similar to helioseismology) 
for the theory of stellar pulsation and the opacity tables at low metallicities. 

\end{document}